\documentclass[amsmath,amssymb,aps,pra,twocolumn,showpacs]{revtex4}
\usepackage{bm}
\usepackage{amsfonts}
\usepackage{amssymb}
\usepackage{graphicx}

\begin{document}

\title{Why photons cannot be sharply localized}
\author{Iwo Bialynicki-Birula}\email{birula@cft.edu.pl}
\affiliation{Center for Theoretical Physics, Polish Academy of Sciences\\
Al. Lotnik\'ow 32/46, 02-668 Warsaw, Poland}
\author{Zofia Bialynicka-Birula}\affiliation{Institute of Physics, Polish Academy of Sciences\\
Al. Lotnik\'ow 32/46, 02-668 Warsaw, Poland}

\begin{abstract}
Photons cannot be localized in a sharply defined region. The expectation value of their energy density and the photon number density can only be approximately localized, leaving an exponential tail. We show that one may sharply localize either electric or magnetic (but not both) footprints of photons, and only momentarily. In the course of time evolution this localization is immediately destroyed. However, the coherent states, like their classical counterparts, can be localized without any limitations. The main tool in our analysis is a set of space-dependent photon creation and annihilation operators defined without any reference to the mode decomposition.
\end{abstract}
\maketitle

\section{Introduction}

Physicists have pondered over the problem of photon localization and the related problem of the photon wave function for almost 80 years now, beginning with the work of Landau and Peierls \cite{lp}. An extensive review of the photon localization problem was recently presented by Keller \cite{ok}.  The problem of photon localization is closely related to the widely studied problem of the photon position operator. In this paper we introduce an operational definition of {\em partial localization}, based on the measurements of correlation functions for electric or magnetic fields. For a want of better names, we shall use the terms {\em electric localization} and {\em magnetic localization} even though this might erroneously suggest the presence of some electric or magnetic devices that confine the photons. Since a sharp localization of photons according to our operational definition of localization is not possible, a photon position operator compatible with this definition does not exist.

Earlier studies of the photon localization emphasized the mathematical aspects (see, for example, \cite{nw,jp,am}). In this paper we emphasize the physical properties of the electromagnetic field. In particular, we exhibit the role of the photon helicity and the symmetry between the electric and magnetic fields. We proceed in the footsteps of Glauber \cite{g,g1,g2,tg} who was the first to recognize the significance of the space-dependent creation and annihilation operators. This approach was recently summarized and expanded in an extensive paper by Smith and Raymer \cite{sr}. In our work we concentrate on the analysis of the photon localization in terms of the electric and magnetic field operators. We treat these fields (after smearing over space-time regions) as {\em bona fide  observables}. We put emphasis on the field aspect that is complementary to the particle aspect. It might have a weaker connection with experiments (usually based on photon counting as highlighted by Glauber) but it is more precise as was explained in detail by Bohr and Rosenfeld \cite{br,br1}

The standard method of quantization of the free electromagnetic field, based on the decomposition into monochromatic modes, is not well suited for the discussion of localizability because the monochromatic mode functions are not localized. To overcome this problem we further developed an alternative method of quantization that does not require a mode decomposition. The essential mathematical tools in our analysis are the Riemann-Silberstein (RS) vector and the helicity operator. This formulation has some merits of its own and it can also be used to study other general properties of photons (for example, the local aspects of entanglement) without restrictions due to a specific choice of modes.

Photons are quanta of the electromagnetic field --- they are the carriers of the electromagnetic field hence every photon has its electric and magnetic side. The localization of photons studied in this paper relies on the electric or magnetic manifestations of the photon's existence --- on the footprints that a photon leaves. We say that the photons are electrically or magnetically localized in a region $R$ if their electric or magnetic properties are confined to the localization region. The electric localization is complementary to the magnetic localization --- electrically localized states are magnetically delocalized and {\em vice versa}. Since the electric and magnetic field vectors are canonically conjugate variables, this is akin to the complementarity of position and momentum in quantum mechanics. States of particles sharply localized in position are spread out in momentum and vice versa.

The localization of photons discussed in this paper is not easily seen in the standard formalism since the mode functions that are introduced there correspond to well defined frequencies and one needs their superpositions to produce localized states. A much more convenient tool is a set of space-dependent creation and annihilation operators discussed in Sec. \ref{cran}. These mode-independent operators enable us to introduce in Sec. \ref{fock} the Fock bases which give a precise meaning to photon states. In Sec. \ref{elmag} we introduce the notion of electrically or magnetically localized photon states and in Sec. \ref{time} we show that their localization is an ephemeral effect --- it is immediately destroyed by the time evolution. A weaker form of localization: diffuse localization is discussed in Sec. \ref{dif} and finally in Sec. \ref{coher} we analyze localization in terms of coherent states.

Throughout this paper we shall work in the Heisenberg picture. The observables will be represented by time-dependent operators while the state vectors will be constructed at a reference time, taken as usual to be $t=0$. Our definition of the creation and annihilation operators {\em does not use} the Hamiltonian or the time-dependence of the field operators. We do not split the field operators into its creation and annihilation parts according to the sign of the frequency. Instead, we rely solely on the equal-time canonical commutation relations.

\section{RS vector and the photon helicity operator}\label{rs0}

One hundred years ago Silberstein \cite{ls,pwf0,pwf,arx,qed,bb} discovered that the following complex combination of the electromagnetic field vectors,
\begin{align}\label{rs}
{\bm F}({\bm r},t)=\frac{{\bm D}({\bm r},t)}{\sqrt{2\varepsilon}}+i\frac{{\bm B}({\bm r},t)}{\sqrt{2\mu}}
\end{align}
is very useful in the analysis of solutions of the Maxwell equations. Indeed, the two pairs of Maxwell equations become one pair without any loss of information
\begin{align}\label{max}
\partial_t{\bm F}({\bm r},t)=-ic{\bm\nabla}\times{\bm F}({\bm r},t),\quad {\bm\nabla}\!\cdot\!{\bm F}({\bm r},t)=0.
\end{align}
One can always recover the ${\bm D}$ and ${\bm B}$ fields by taking the real and imaginary part, but treating ${\bm F}(\bm r,t)$ as one object simplifies the analysis. In quantum electrodynamics the classical function ${\bm F}(\bm r,t)$ is replaced by the operator ${\hat{\bm F}}(\bm r,t)$ in the Heisenberg picture satisfying the Maxwell equations.

In particle physics the helicity operator ${\hat\chi}$ is defined as the projection ${\bm s}\cdot{\bm n}$ of the spin operator  ${\bm s}$ on the direction of momentum  ${\bm n}={\bm k}/k$. The helicity of photons, and other massless particles, has a Lorentz-invariant meaning. For spin-one particles the spin operator is a vector whose components are $3\times 3$ matrices. The components of these matrices can be expressed in terms of the Levi-Civita symbol in the following way (in units of $\hbar$):
\begin{eqnarray}
\{s_i\}_{jk}=-i\epsilon_{ijk}.
\end{eqnarray}
Thus, the action of the helicity operator on any vector function in momentum space is represented by the following matrix:
\begin{eqnarray}
{\hat\chi}={\bm n}\!\cdot\!{\bm s} = \left(\begin{array}{ccc}
0 & -in_z & in_y\\
in_z & 0 & -in_x\\
-in_y & in_x & 0\end{array}
\right),
\end{eqnarray}
and is equivalent to the following cross product :
\begin{align}\label{cp}
{\hat\chi}=\frac{i{\bm k}\times}{k}.
\end{align}
For plane waves, the positive (negative) helicity corresponds to the left-handed (right-handed) circular polarization.

In position space, the cross product with the vector $i{\bm k}$ becomes the curl and the division by $k$ is represented by
\begin{align}\label{nonl}
\int\!\frac{d^3k}{(2\pi)^3}\frac{e^{i{\bm k}\cdot({\bm r-\bm r'})}}{k}=\frac{1}{2\pi^2|\bm r-\bm r'|^2}.
\end{align}
Therefore, the action of the helicity operator ${\hat\chi}$ on any vector function ${\bm V}(\bm r)$ becomes the following nonlocal operation:
\begin{align}\label{chi}
{\hat\chi}{\bm V}(\bm r)=\frac{1}{2\pi^2}\int\!d^3r'\frac{1}{|\bm r-\bm r'|^2}{\bm\nabla}\times{\bm V}(\bm r').
\end{align}
In the subspace of divergence-free vectors, the helicity operator is an involution, ${\hat\chi}^2=1$. This property implies that the operators $P_\pm$,
\begin{align}\label{proj}
P_\pm=\frac{1\pm{\hat\chi}}{2},\quad P_\pm^2=P_\pm,\quad P_+P_-=0,
\end{align}
are projectors on the positive and negative helicity subspaces. In Sec.~\ref{cran} we show that with the use of the helicity operator we can split the operator ${\hat{\bm F}}(\bm r,t)$ into its creation and annihilation parts.

\section{Space-dependent creation and annihilation operators}\label{cran}

The components of the RS operator ${\hat{\bm F}}({\bm r},t)$ and its Hermitian conjugate satisfy the equal-time commutation relations (repeated indices always imply summation)
\begin{align}\label{cr}
\left[{\hat F}_i({\bm r},t),{\hat F}^\dagger_j({\bm r}',t)\right]=\hbar c\,\epsilon_{ilj}\partial_l\delta^{(3)}({\bm r}-{\bm r}'),
\end{align}
that follow directly from the canonical commutation relations \cite{pj,hp} for the electromagnetic field operators ${\hat{\bm D}}({\bm r},t)$ and ${\hat{\bm B}}({\bm r},t)$,
\begin{align}\label{ccr}
\left[{\hat B}_i({\bm r},t),{\hat D}_j({\bm r}',t)\right]=-i\hbar\,\epsilon_{ilj}\partial_l\delta^{(3)}({\bm r}-{\bm r}').
\end{align}
By acting on the RS operator with the projection operators [(\ref{proj})],
\begin{align}\label{fpm}
{\hat{\bm F}}^\pm=P_\pm{\hat{\bm F}},
\end{align}
we split ${\hat{\bm F}}({\bm r},t)$ into its annihilation, ${\hat{\bm F}}^+$, and creation, ${\hat{\bm F}}^-$, parts. This identification follows from the commutation relations for these two parts of the RS operator (the remaining commutators vanish):
\begin{subequations}\label{crpm}
\begin{align}
\left[{\hat F}_i^+({\bm r},t),{\hat F}^{+\dagger}_j({\bm r}',t)\right]&=\hbar c\,c_{ij}^+({\bm r}-{\bm r}'),\\
\left[{\hat F}_i^-({\bm r},t),{\hat F}^{-\dagger}_j({\bm r}',t)\right]&=-\hbar c\,c_{ij}^-({\bm r}-{\bm r}'),
\end{align}
\end{subequations}
where the functions $c_{ij}^\pm$ are
\begin{align}\label{cpm}
&c_{ij}^\pm({\bm r}-{\bm r}')\\
&=\frac{1}{4\pi^2}\left(\partial_i\partial_j-\delta_{ij}\Delta\right)\frac{1}{|{\bm r}-{\bm r}'|^2}\pm\frac{1}{2}\epsilon_{ilj}\partial_l\delta^{(3)}({\bm r}-{\bm r}').\nonumber
\end{align}
These relations are obtained from the definition of the projector operators [(\ref{proj})] and from the commutation relations [(\ref{cr})]. The significance of this result is best seen in Fourier space,
\begin{align}\label{four}
{\tilde c}_{ij}^\pm({\bm k})=\frac{\delta_{ij}k^2-k_ik_j}{2k}\pm\frac{i}{2}\epsilon_{ilj}k_l.
\end{align}
Both Hermitian matrices ${\tilde c}_{ij}^+({\bm k})$ and ${\tilde c}_{ij}^-({\bm k})$ have the eigenvalues $k$ and $0$. The minus sign in the second commutation relation [(\ref{crpm})] means that the operator ${\hat{\bm F}}^-$ must be interpreted as a creation operator. In accordance with this interpretation, we introduce the following notation:
\begin{subequations}\label{cran1}
\begin{align}
{\bm d}({\bm r},t)&=\frac{{\hat{\bm F}}^+({\bm r},t)+{\hat{\bm F}}^{-\dagger}({\bm r},t)}{\sqrt{\hbar c}},\label{cran11}\\
{\bm d}^\dagger({\bm r},t)&=\frac{{\hat{\bm F}}^-({\bm r},t)+{\hat{\bm F}}^{+\dagger}({\bm r},t)}{\sqrt{\hbar c}},\\
{\bm b}({\bm r},t)&=\frac{{\hat{\bm F}}^+({\bm r},t)-{\hat{\bm F}}^{-\dagger}({\bm r},t)}{i\sqrt{\hbar c}},\label{cran13}\\
{\bm b}^\dagger({\bm r},t)&=\frac{{\hat{\bm F}}^-({\bm r},t)-{\hat{\bm F}}^{+\dagger}({\bm r},t)}{i\sqrt{\hbar c}}.
\end{align}
\end{subequations}
It follows from their definition that the operators ${\bm d}$ and ${\bm b}$ are not independent,
\begin{subequations}\label{db}
\begin{align}
{\bm b}({\bm r},t)&=-i{\hat\chi}{\bm d}({\bm r},t),\\
{\bm b}^\dagger({\bm r},t)&=i{\hat\chi}{\bm d}^\dagger({\bm r},t).
\end{align}
\end{subequations}
All four vector operators ${\bm d}$, ${\bm d}^\dagger$, ${\bm b}$, and ${\bm b}^\dagger$ are divergence-free. Their commutation relations are
\begin{subequations}\label{cra}
\begin{align}
\left[d_i({\bm r},t),d_j^\dagger({\bm r}',t)\right]&=(\partial_i\partial_j-\delta_{ij}\Delta)
\frac{1}{2\pi^2|{\bm r}-{\bm r}'|^2},\\
\left[b_i({\bm r},t),b_j^\dagger({\bm r}',t)\right]&=(\partial_i\partial_j-\delta_{ij}\Delta)
\frac{1}{2\pi^2|{\bm r}-{\bm r}'|^2},\\
\left[b_i({\bm r},t),d_j^\dagger({\bm r}',t)\right]&=-i\epsilon_{ilj}\partial_l\delta^{(3)}({\bm r}-{\bm r}').\label{cra3}
\end{align}
\end{subequations}
The operators [(\ref{cran1})] represent the annihilation and creation parts of the field operators,
\begin{subequations}\label{cran3}
\begin{align}
{\hat{\bm D}}({\bm r},t)&=\sqrt{\frac{\hbar c\varepsilon}{2}}\left[{\bm d}^\dagger({\bm r},t)+{\bm d}({\bm r},t)\right],\\
{\hat{\bm B}}({\bm r},t)&=\sqrt{\frac{\hbar c\mu}{2}}\left[{\bm b}^\dagger({\bm r},t)+{\bm b}({\bm r},t)\right].
\end{align}
\end{subequations}
The roles of these annihilation and creation operators will be fully elucidated in Sec.~\ref{fock} where we construct the Fock bases.

\section{Photon number operator and the Fock bases}\label{fock}

We shall use the expression for the operator of the total number of photons, given by Zeldovich \cite{yz}. It has the following nonlocal form:
\begin{align}\label{n1}
{\hat N}&=\frac{1}{4\pi^2\hbar c}\int\!d^3r\int\!d^3r'\nonumber\\
&\times :\!\left[\frac{{\hat{\bm D}}({\bm r},t)\!\cdot\!{\hat{\bm D}}({\bm r}',t)}{\varepsilon|{\bm r}-{\bm r}'|^2}
+\frac{{\hat{\bm B}}({\bm r},t)\!\cdot\!{\hat{\bm B}}({\bm r}',t)}{\mu|{\bm r}-{\bm r}'|^2}\right]\!:.
\end{align}
We have omitted the time argument of ${\hat N}$ because the photon number operator is time-independent, as a result of Maxwell equations. With the use of Eqs.~(\ref{cran3}), we can explicitly carry out the normal ordering and obtain,
\begin{align}\label{n}
{\hat N}=\int\!\!d^3r\!\int\!\!d^3r'\frac{{\bm d}^\dagger({\bm r},t)\!\cdot\!{\bm d}({\bm r}',t)+{\bm b}^\dagger({\bm r},t)\!\cdot\!{\bm b}({\bm r}',t)}{4\pi^2|{\bm r}-{\bm r}'|^2}.
\end{align}
One can prove, using Eqs.~(\ref{db}), that the contributions from the electric and magnetic parts to the total number of photons are equal. Using formula (\ref{a2}) one can transform the number operator to the form,
\begin{align}\label{cook}
{\hat N}=\int\!d^3r\left[{\hat{\bm\psi}}^\dagger({\bm r},t)\!\cdot\!{\hat{\bm\psi}}({\bm r},t)+{\hat{\bm\phi}}^\dagger({\bm r},t)\!\cdot\!{\hat{\bm\phi}}({\bm r},t)\right],
\end{align}
where the operators ${\hat{\bm\psi}}$ and ${\hat{\bm\phi}}$ introduced by Cook \cite{cook} are connected with ${\bm d}$ and ${\bm b}$ through the equations
\begin{subequations}\label{cook1}
\begin{align}
{\hat{\bm\psi}}({\bm r},t)=\frac{1}{8\pi^{3/2}}\int\frac{d^3r'}{|{\bm r}-{\bm r}'|^{5/2}}{\bm d}({\bm r}',t),\\
{\hat{\bm\phi}}({\bm r},t)=\frac{1}{8\pi^{3/2}}\int\frac{d^3r'}{|{\bm r}-{\bm r}'|^{5/2}}{\bm b}({\bm r}',t).
\end{align}
\end{subequations}

The creation and annihilation operators satisfy the standard commutation relations with the photon number operator;
\begin{subequations}\label{ncra}
\begin{align}
\left[{\hat N},{\bm d}^\dagger({\bm r},t)\right]&={\bm d}^\dagger({\bm r},t),\;\;\left[{\hat N},{\bm d}({\bm r},t)\right]=-{\bm d}({\bm r},t),\\
\left[{\hat N},{\bm b}^\dagger({\bm r},t)\right]&={\bm b}^\dagger({\bm r},t),\;\;\left[{\hat N},{\bm b}({\bm r},t)\right]=-{\bm b}({\bm r},t),
\end{align}
\end{subequations}
obtained with the use of the commutation relations (\ref{cra}), Eq.~(\ref{a1}), and the formula
\begin{align}\label{lap}
\Delta\frac{1}{|{\bm r}'-{\bm r}''|}=-4\pi\delta^{(3)}({\bm r}'-{\bm r}'').
\end{align}
As was to be expected, the creation operator increases and the annihilation operator decreases the number of photons by one.

The photon creation and annihilation operators ${\bm d}^\dagger({\bm r},t),{\bm b}^\dagger({\bm r},t),{\bm d}({\bm r},t)$, and ${\bm b}({\bm r},t)$ differ from their standard counterparts: they are not labeled by quantum numbers (such as momentum and angular momentum) or by the mode number $l$ enumerating some wave-packet modes ${\bm v}_l({\bm r})$ as in \cite{tg,sr}. Instead, they are labeled by space points and they also have a vector character. We emphasize that in their definition we have neither used a mode decomposition nor the equations of motion.

Having established the basic properties, we can proceed further and construct the Fock basis. To this end we choose $t=0$ as our reference time for the construction of state vectors in the Heisenberg picture, and we drop the time parameter. We can construct the Fock basis with the use of either ${\bm d}^\dagger({\bm r})$ or ${\bm b}^\dagger({\bm r})$ operator. We begin with the operators ${\bm d}^\dagger({\bm r})$. They furnish the space of states with the structure of an inverted pyramid of Fock states
\begin{align}
\ldots\ldots\ldots\ldots\ldots\ldots\ldots\ldots\ldots\ldots\ldots\ldots\ldots\nonumber\\
d_i^\dagger({\bm r})d_j^\dagger({\bm r}\,')d_k^\dagger({\bm r}\,'')d_l^\dagger({\bm r}\,''')|0\rangle\quad n=4\nonumber\\
d_i^\dagger({\bm r})d_j^\dagger({\bm r}\,')d_k^\dagger({\bm r}\,'')|0\rangle\quad n=3\nonumber\\
d_i^\dagger({\bm r})d_j^\dagger({\bm r}\,')|0\rangle\quad n=2\nonumber\\
d_i^\dagger({\bm r})|0\rangle\quad n=1\nonumber\\
|0\rangle\quad n=0
\end{align}
These state vectors form a complete basis --- every state can be described by the superposition of the basis vectors. Thus, an arbitrary $n$-photon state has the form
\begin{align}\label{aa}
\int\!d^3r_1\dots\int\!d^3r_n\varphi^{i_1\dots i_n}({\bm r}_1,\dots {\bm r}_n)
d^\dagger_{i_1}({\bm r}_1)\dots d^\dagger_{i_n}({\bm r}_n)|0\rangle.
\end{align}
This basis is not orthogonal due to nonlocal commutation relations between the operators ${\bm d}({\bm r})$ and ${\bm d}^\dagger({\bm r}')$. The second Fock basis can be constructed with the use of the operators ${\bm b}^\dagger({\bm r})$. These two bases are dual  \cite{db} with respect to each other (their vectors are mutually orthogonal) because the commutation relations between ${\bm b}({\bm r})$ and ${\bm d}^\dagger({\bm r}')$ are local. In particular, the orthogonality condition for one-photon basis vectors has the form:
\begin{align}\label{db1}
\langle 0|b_i({\bm r})d^\dagger_j({\bm r}')|0\rangle=-i\epsilon_{ilj}\partial_l\delta^{(3)}({\bm r}-{\bm r}').
\end{align}
The orthogonality for $n$-photon state vectors also follows from commutation relation (\ref{cra3}).

\section{Electric and magnetic photon wave functions}\label{wf}

Every one-photon state can be obtained by acting on the vacuum state $|0\rangle$ with the creation operator $a^\dagger$ constructed from either ${\bm d}^\dagger$ or ${\bm b}^\dagger$,
\begin{align}\label{a}
a^\dagger=\int\!d^3r\,{\bm\varphi}({\bm r})\!\cdot\!{\bm d}^\dagger({\bm r})=\int\!d^3r\,{\tilde{\bm\varphi}}({\bm r})\!\cdot\!{\bm b}^\dagger({\bm r}),
\end{align}
where ${\bm\varphi}({\bm r})$ and ${\tilde{\bm\varphi}}({\bm r})$ are complex-valued vector wave functions. To guarantee the equality of both integrals, these two wave functions must be connected by the relation
\begin{align}\label{by}
{\tilde{\bm\varphi}}({\bm r})=-i{\hat\chi}{\bm\varphi}({\bm r}).
\end{align}
We shall assume throughout this paper that all vector wave functions are divergence-free. The normalization condition for both these functions has a nonlocal form,
\begin{align}\label{norm}
\frac{1}{2\pi^2}\int\!d^3r\int\!d^3r'\!\frac{\left[{\bm\nabla}\times{\bm\varphi}^*({\bm r})\right]\!\cdot\!\left[{\bm\nabla}\times{\bm\varphi}({\bm r}')\right]}{|{\bm r}-{\bm r}'|^2}=1,
\end{align}
because of the nonlocal form of commutation relations (\ref{cra}). This condition guarantees the standard normalization of the commutator $[a,a^\dagger]=1$.

Even though both wave functions in Eqs.~(\ref{a}) can be used equally well to describe every photon state, one may be preferred over the other depending on the situation. It is natural to call ${\bm\varphi}({\bm r})$ as the {\em electric} wave function and ${\tilde{\bm\varphi}}({\bm r})$ as the {\em magnetic} wave function. Every particular wave function can be used in two different ways: it can be treated as an electric or magnetic wave function, leading to two {\em different} photon states. These two states differ by an interchange of their electric and magnetic properties (duality transformation). The existence of two photon states described by the same function reflects perfect symmetry between electricity and magnetism of the free electromagnetic radiation field.

The functions ${\bm\varphi}({\bm r})$ and ${\tilde{\bm\varphi}}({\bm r})$ do not coincide with the photon wave functions studied before \cite{pwf0,pwf,arx,sr} but they are directly related. The photon wave functions ${\bm{\mathcal F}}_\pm$ for both signs of helicity were defined in \cite{pwf} as the matrix elements of the RS operator and its Hermitian conjugate taken between the vacuum and an arbitrary one-photon state vector,
\begin{subequations}\label{rswf}
\begin{align}
{\bm{\mathcal F}}_+({\bm r},t)=\langle 0|{\hat{\bm F}}({\bm r},t)a^\dagger|0\rangle,\\
{\bm{\mathcal F}}_-({\bm r},t)=\langle 0|{\hat{\bm F}}^\dagger({\bm r},t)a^\dagger|0\rangle.
\end{align}
\end{subequations}
By adding and subtracting these two equations, with the use of Eqs.~(\ref{cran3}), (\ref{db1}), and (\ref{a}), we obtain at $t=0$
\begin{subequations}\label{wfem}
\begin{align}
\frac{{\bm{\mathcal F}}_+({\bm r})+{\bm{\mathcal F}}_-({\bm r})}{\sqrt{\hbar c}}&=\int\!d^3r'\langle 0|{\bm d}({\bm r}){\tilde{\bm\varphi}}({\bm r}')\!\cdot\!{\bm b}^\dagger({\bm r}')|0\rangle,\nonumber\\
&=-i{\bm\nabla}\times{\tilde{\bm\varphi}}({\bm r}),\\
\frac{{\bm{\mathcal F}}_+({\bm r})-{\bm{\mathcal F}}_-({\bm r})}{\sqrt{\hbar c}}&=\int\!d^3r'\langle 0|{\bm b}({\bm r},0){\bm\varphi}({\bm r}')\!\cdot\!{\bm d}^\dagger({\bm r}')|0\rangle,\nonumber\\
&={\bm\nabla}\times{\bm\varphi}({\bm r}).
\end{align}
\end{subequations}
These relations between both types of photon wave functions can be extended to all $t$ because the wave functions obey the same evolution equations obtained from the Maxwell equations.

\section{Electric and magnetic localizations}\label{elmag}

We adopt the following operational definition of the electric (magnetic) localizability. A photon state is electrically (magnetically) localized in some region $R$ if all measurements of the electric (magnetic) fields (including their correlations) carried out outside of $R$ do not reveal the existence of the photon.

In order to construct localized photon states we choose a sharply localized (vanishing outside of $R$) vector function ${\bm\psi}_R({\bm r})$. Depending on whether we treat this function as an electric or a magnetic photon wave function, we obtain two different photon states created by the operators, $a^\dagger_R$ and ${\tilde a}^\dagger_R$,
\begin{subequations}\label{ar}
\begin{align}
a^\dagger_R=\int\!d^3r\,{\bm\psi}_R({\bm r})\!\cdot\!{\bm d}^\dagger({\bm r}),\label{ar1}\\
{\tilde a}^\dagger_R=\int\!d^3r\,{\bm\psi}_R({\bm r})\!\cdot\!{\bm b}^\dagger({\bm r}).\label{ar2}
\end{align}
\end{subequations}
They obey the following commutation relations:
\begin{subequations}\label{mcr}
\begin{align}
\left[{\hat{\bm D}}({\bm r},0),a^\dagger_R\right]
&=\sqrt{\frac{\hbar c\varepsilon}{2}}\,{\bm\nabla}\times{\hat\chi}{\bm\psi}_R({\bm r}),\label{mcr1}\\
\left[{\hat{\bm B}}({\bm r},0),a^\dagger_R\right]
&=-i\sqrt{\frac{\hbar c\mu}{2}}\,{\bm\nabla}\times{\bm\psi}_R({\bm r}),\label{mcr2}\\
\left[{\hat{\bm D}}({\bm r},0),{\tilde a}^\dagger_R\right]
&=i\sqrt{\frac{\hbar c\varepsilon}{2}}\,{\bm\nabla}\times{\bm\psi}_R({\bm r}),\label{mcr3}\\
\left[{\hat{\bm B}}({\bm r},0),{\tilde a}^\dagger_R\right]
&=\sqrt{\frac{\hbar c\mu}{2}}\,{\bm\nabla}\times{\hat\chi}{\bm\psi}_R({\bm r}).\label{mcr4}
\end{align}
\end{subequations}
We shall show that these two operators create either magnetically or electrically localized states. Indeed, when ${\bm r}$ lies outside of $R$, the operator $a^\dagger_R$ commutes with the operator ${\hat{\bm B}}({\bm r},0)$. This property guarantees the magnetic localizability. To show this, let us consider the results of the measurements of the correlation functions of the magnetic field. They are described by the expectation values of any number of the operators of the magnetic field,
\begin{align}\label{exp}
\langle 0|a_R{\hat B}_i({\bm r}_1,0){\hat B}_j({\bm r}_2,0)\dots{\hat B}_k({\bm r}_n,0)a^\dagger_R|0\rangle.
\end{align}
When all points ${\bm r}_i$ lie outside of $R$, the operator $a^\dagger_R$ commutes with all field operators ${\hat B}_i({\bm r}_i)$. Therefore, we may bring ${\hat a}^\dagger_R$ all the way to the left, and with the use of normalization condition (\ref{norm}) we obtain the expectation value in the vacuum state, as if the photon was not present, namely,
\begin{align}\label{res}
&\langle 0|a_R{\hat B}_i({\bm r}_1,0){\hat B}_j({\bm r}_2,0)\dots{\hat B}_k({\bm r}_n,0)a^\dagger_R|0\rangle\nonumber\\
&=\langle 0|{\hat B}_i({\bm r}_1,0){\hat B}_j({\bm r}_2,0)\dots{\hat B}_k({\bm r}_n,0)|0\rangle.
\end{align}
According to Eq.~(\ref{mcr1}) the correlation functions of the electric field involve the function ${\hat\chi}{\bm\psi}_R({\bm r})$. We prove in Appendix \ref{aloc} that this function {\em is not localized}. Therefore, the expectation values of the products of electric field vectors in the magnetically localized photon state differ from the vacuum values outside of the localization region --- the electric field is felt throughout all space. One may say, therefore, that in such a state the magnetic properties are hidden inside of $R$ while the electric properties are spread throughout the whole space.

Creation operator (\ref{ar2}) can also be chosen in the form:
\begin{align}\label{art}
{\tilde a}^\dagger_R=i\int\!d^3r\,{\hat\chi}{\bm\psi}_R({\bm r})\!\cdot\!{\bm d}^\dagger({\bm r}).
\end{align}
The one-photon state created by ${\tilde a}^\dagger_R$ is magnetically delocalized but it is electrically localized,
\begin{align}\label{res1}
&\langle 0|{\tilde a}_R{\hat D}_i({\bm r}_1,0){\hat D}_j({\bm r}_2,0)\dots{\hat D}_k({\bm r}_n,0){\tilde a}^\dagger_R|0\rangle\nonumber\\
&=\langle 0|{\hat D}_i({\bm r}_1,0){\hat D}_j({\bm r}_2,0)\dots{\hat D}_k({\bm r}_n,0)|0\rangle.
\end{align}

The energy operator ${\hat H}$ of the electromagnetic field, unlike the photon number operator [(\ref{n})], is given by a local expression
\begin{align}
{\hat H}&=\int\!d^3r:\!\left[\frac{{\hat{\bm D}}({\bm r},t)\!\cdot\!{\hat{\bm D}}({\bm r},t)}{2\varepsilon}
+\frac{{\hat{\bm B}}({\bm r},t)\!\cdot\!{\hat{\bm B}}({\bm r},t)}{2\mu}\right]\!:\label{h}\\
&=\frac{1}{2}\hbar c\int\!d^3r\left[{\bm d}^\dagger({\bm r},t)\!\cdot\!{\bm d}({\bm r},t)+{\bm b}^\dagger({\bm r},t)\!\cdot\!{\bm b}({\bm r},t)\right].\label{h1}
\end{align}
The two terms in the last integrand are not equal to the electric and magnetic energy densities that appear in Eq.~(\ref{h}) because they do not contain the nondiagonal terms that change the total number of photons. These terms canceled out only due to the integration over all space. However, in a restricted sense the operators
\begin{subequations}\label{ed}
\begin{align}
{\hat{\cal H}}_e({\bm r},t)={\bm d}^\dagger({\bm r},t)\!\cdot\!{\bm d}({\bm r},t),\\
{\hat{\cal H}}_m({\bm r},t)={\bm b}^\dagger({\bm r},t)\!\cdot\!{\bm b}({\bm r},t),
\end{align}
\end{subequations}
can be interpreted as the {\em photon} electric and magnetic energy densities because the nondiagonal terms do not contribute to the expectation values in $n$-photon states. Note that the total magnetic and electric energies are equal.

The expectation values of ${\hat{\cal H}}_e$ and ${\hat{\cal H}}_m$ at $t=0$ in a one-photon state can be calculated with the use of commutation relations (\ref{cra}). Due to the complementarity between the electric localization and magnetic localization, the full energy density ${\hat{\cal H}}_e+{\hat{\cal H}}_m$ in a one-photon state cannot be sharply localized. One can only localize separately either the electric or the magnetic part.

The same properties hold for the photon number density operator (\ref{cook}). In order to obtain the magnetic localization we choose the photon state created by operator [(\ref{a})] with the wave function ${\bm\varphi}$ in the form
\begin{align}\label{nloc}
{\bm\varphi}({\bm r})={\bm\nabla}\times\frac{1}{2\pi^{3/2}}\int\frac{d^3r'}{|{\bm r}-{\bm r}'|^{3/2}}{\bm\psi}_R({\bm r}').
\end{align}
The expectation value of the magnetic part of the photon number density is then localized in $R$,
\begin{align}
\langle 0|a\,{\hat{\bm\phi}}^\dagger({\bm r},t)\!\cdot\!{\hat{\bm\phi}}({\bm r},t)\,a^\dagger|0\rangle
=|{\bm\psi}_R({\bm r})|^2.
\end{align}
However, owing again to the presence of the helicity operator, the electric part is delocalized,
\begin{align}
\langle 0|a\,{\hat{\bm\psi}}^\dagger({\bm r},t)\!\cdot\!{\hat{\bm\psi}}({\bm r},t)\,a^\dagger|0\rangle
=|{\hat{\chi}}{\bm\psi}_R({\bm r})|^2.
\end{align}
Analogous results are obtained, {\em mutatis mutandis}, for the electric part of the photon number density that can be localized while the magnetic part is delocalized.

\section{Destruction of photon localization by time evolution} \label{time}

Up to this point we have not used the Maxwell equations or any other elements of the photon dynamics. All conclusions were based on the instantaneous properties of the system. The impossibility of a {\em simultaneous} electric and magnetic localizations follows directly from the complementarity between electric and magnetic nature of a photon.

Even this partial (electric or magnetic) localization can be achieved {\em only momentarily}. It is immediately destroyed by the time evolution. The solution of the initial value problem for the Maxwell equations (see, for example, \cite{qed}) shows that the electric and magnetic field operators at $t\neq 0$ depend on {\em both} field vectors at $t=0$,
\begin{subequations}\label{ivp}
\begin{align}
{\hat{\bm D}}({\bm r},t)&=\int\!d^3r'\big[\partial_t D({\bm r}-{\bm r}',t){\hat{\bm D}}({\bm r}',0)\nonumber\\
&+D({\bm r}-{\bm r}',t){\bm\nabla}\times{\hat{\bm B}}({\bm r}',0)/\mu\big],\\
{\hat{\bm B}}({\bm r},t)&=\int\!d^3r'\big[\partial_t D({\bm r}-{\bm r}',t){\hat{\bm B}}({\bm r}',0)\nonumber\\
&-D({\bm r}-{\bm r}',t){\bm\nabla}\times{\hat{\bm D}}({\bm r}',0)/\varepsilon\big],
\end{align}
\end{subequations}
where $D({\bm r},t)$ is the Jordan-Pauli function \cite{pj},
\begin{align}\label{jp}
D({\bm r},t)=\frac{1}{4\pi cr}\left[\delta(ct-r)-\delta(ct+r)\right].
\end{align}
Thus, the expectation values of the magnetic fields at $t$,
\begin{align}\label{expt}
\langle 0|a_R{\hat B}_i({\bm r}_1,t){\hat B}_j({\bm r}_2,t)\dots{\hat B}_k({\bm r}_n,t)a^\dagger_R|0\rangle,
\end{align}
are coupled to the expectation values involving {\em both} the magnetic and the electric fields at $t=0$. Since a magnetically localized photon is not electrically localized, even its partial localization is an ephemeral effect.

\section{Diffuse localization of photon states}\label{dif}

Since we have shown that sharp localization in a restricted region is impossible, we can only try to localize photons approximately. It turns out that we can construct photon states \cite{loc} for which the energy density has an exponential tail with an arbitrarily small scale factor $l$. The simplest example of such a state is characterized by the square-root exponential fall-off. Stronger localization is also possible but then we have to resort to special functions. Let us choose the wave function in the form \cite{loc}
\begin{align}\label{exp1}
{\bm\psi}_D({\bm r})={\bm m}\,{\rm Im}\frac{e^{-2\sqrt{1-ir/l}}}{r}={\bm m}\frac{e^{-\kappa_+}}{r}\sin\kappa_-,
\end{align}
where ${\bm m}$ is a constant vector that includes also the normalization factor and
\begin{align}\label{s}
\kappa_\pm=\sqrt{2}\sqrt{\sqrt{1+(r/l)^2}\pm 1}.
\end{align}
With the help of the projector [(\ref{proj})] we can construct the eigenfunction ${\bm\psi}_H({\bm r})=P_+{\bm\psi}_D({\bm r})$ of the helicity operator. In Appendix \ref{adif} we evaluate ${\hat\chi}{\bm\psi}_D$ and obtain a function with the same exponential fall-off as ${\bm\psi}_D$,
\begin{align}\label{exp2}
&{\hat\chi}{\bm\psi}_D({\bm r})=\frac{{\bm r}\times{\bm m}}{r}\frac{e^{-\kappa_+}}{r}\nonumber\\
&\times\left[\left(1+\frac{l}{2r}\kappa_+\right)\cos\kappa_--\frac{l}{2r}\left(1+\kappa_+\right)\sin\kappa_-\right].
\end{align}
The creation operator $a_H^\dagger$ constructed with the use of the function ${\bm\psi}_H$ will have the same commutation relations (up to factors $\pm i$) with both electric and magnetic fields. In this case the electric and magnetic localizations will be governed by the same law: the expectation values of all the field vectors, electric and magnetic, will fall-off as $e^{-\sqrt{2r/l}}/r$ when one moves away from the localization center. Therefore, we can choose the scale factor $l$ so that at a given distance {\em all field correlation functions} will be arbitrarily small. Of course, these functions will spread out with the speed of light during time evolution \cite{loc}. Diffuse localization with an exponential tail also holds for the photon number density.

\section{Sharp localization of coherent states}\label{coher}

All the limitations of the localization apply to the states with a given number of photons. Coherent states, as we shall show now, are not subject to such limitations. The easiest way to see this employs the displacement operator \cite{g}. The displacement operator is usually constructed from the creation and annihilation operators that result from a mode decomposition. Here we shall construct the displacement operator from our space-dependent operators. The most general coherent state can be written in two equivalent forms
\begin{align}\label{coh}
&\exp\int\!d^3r\,\left[{\bm\varphi}({\bm r})\!\cdot\!{\bm d}^\dagger({\bm r})
-{\bm\varphi}^*({\bm r})\!\cdot\!{\bm d}({\bm r})\right]|0\rangle\nonumber\\
&=\exp\int\!d^3r\,\left[{\tilde{\bm\varphi}}({\bm r})\!\cdot\!{\bm b}^\dagger({\bm r})
-{\tilde{\bm\varphi}}^*({\bm r})\!\cdot\!{\bm b}({\bm r})\right]|0\rangle.
\end{align}
This formula expressed in terms of the field operators is
\begin{align}\label{coh1}
|{\bm\alpha},{\bm\beta}\rangle=\exp\,i\!\!\int\!d^3r\!\left[{\bm\alpha}({\bm r})\!\cdot\!{\hat{\bm D}}({\bm r})
+{\bm\beta}({\bm r})\!\cdot\!{\hat{\bm B}}({\bm r})\right]|0\rangle.
\end{align}
The two real vector functions ${\bm\alpha}({\bm r})$ and ${\bm\beta}({\bm r})$ are certain combinations of the original complex functions ${\bm\varphi}({\bm r})$ or ${\tilde{\bm\varphi}}({\bm r})$, but at this point we treat them as the primary objects. In order to obtain a sharply localized coherent state we assume that ${\bm\alpha}={\bm\alpha}_R$ and ${\bm\beta}={\bm\beta}_R$ vanish outside of some region $R$. The expectation values of any number of field operators in such a localized state are
\begin{align}\label{coh2}
\langle{\bm\alpha}_R,{\bm\beta}_R|{\hat C}_i({\bm r}_1,0){\hat C}_j({\bm r}_2,0)\dots{\hat C}_k({\bm r}_n,0)|{\bm\alpha}_R,{\bm\beta}_R\rangle,
\end{align}
where the operators ${\hat C}$ stand for ${\hat D}$ or ${\hat B}$. When all points ${\bm r}_i$ lie outside of $R$, we can freely move the operator appearing in Eq.~(\ref{coh1}) to the left since the field operators commute for space-like separations. Next, we use the unitarity of displacement operator (\ref{coh1}) to convert Eq.~(\ref{coh2}) to the vacuum expectation value,
\begin{align}\label{coh3}
\langle 0|{\hat C}_i({\bm r}_1,0){\hat C}_j({\bm r}_2,0)\dots{\hat C}_k({\bm r}_n,0)|0\rangle.
\end{align}
Thus, no measurements in the sense of \cite{br,br1} of the electric and magnetic fields carried out outside of $R$ will reveal the presence of the coherent state localized in $R$. Coherent states may be sharply localized in an arbitrarily small region. The localization region remains sharp but it will expand in time with the speed of light, as seen in formula (\ref{ivp}).

The localization of the field operators implies the localization of the energy density. One can also obtain the full localization of the photon number density by choosing the functions ${\bm\alpha}({\bm r})$ and ${\bm\beta}({\bm r})$ in form (\ref{nloc}).

\section{Conclusions}

We have shown that the symmetry between electricity and magnetism permeating the classical and quantum theories of the electromagnetic field enables us to split the notion of photon localizability into two separate notions: the electric and magnetic localizabilities. We defined operationally the electric (or magnetic) localization of photon states as follows: the electric (or magnetic) footprints of electrically (or magnetically) localized photons are not detectable outside of the localization region. Since the electric and magnetic fields are complementary, electric and magnetic localizabilities are mutually exclusive. We have further shown that even this partial sharp localization of a photon can be achieved only at an instant. This localization is fragile, it is immediately destroyed by time evolution. Due to lack of a sharp photon localization, we constructed diffuse photon states that are simultaneously {\em electrically and magnetically} localized with finite but arbitrary accuracy. We have also shown that there are no restrictions on the localization of coherent states.

\acknowledgments

We would like to thank Marlan Scully, who many years ago inspired us to think about the photon wave function. We are also indebted to Michael Raymer for his thoughtful comments and helpful suggestions. This work was partly supported by the grant from the Polish Ministry of Science and Higher Education.

\appendix
\section{}\label{eqs}

In our calculations we need the following special cases:
\begin{align}\label{a1}
\int\!\frac{d^3r}{|{\bm r}'-{\bm r}|^2|{\bm r}-{\bm r}''|^2}=\frac{\pi^3}{|{\bm r}'-{\bm r}''|},
\end{align}
\begin{align}\label{a2}
\int\frac{d^3r}{|{\bm r}'-{\bm r}|^{5/2}|{\bm r}-{\bm r}''|^{5/2}}=\frac{16\pi}{|{\bm r}'-{\bm r}''|^2},
\end{align}
\begin{align}\label{a3}
\int\frac{ d^3r}{|{\bm r}'-{\bm r}|^{5/2}|{\bm r}-{\bm r}''|^{3/2}}=\frac{4\pi^2}{|{\bm r}'-{\bm r}''|},
\end{align}
of the general formula
\begin{align}\label{general}
&\int\frac{d^3r}{|{\bm r}'-{\bm r}|^\alpha|{\bm r}-{\bm r}''|^\beta} =\sin(\frac{\pi\alpha}{2}) \sin(\frac{\pi\beta}{2}) \sin\pi(\frac{\alpha+\beta}{2})\nonumber\\
&\quad\quad\quad\times\frac{8\Gamma(\alpha+\beta-4)\Gamma(2-\alpha)\Gamma(2-\beta)}{|{\bm r}'-{\bm r}''|^{\alpha+\beta-3}}.
\end{align}

\section{}\label{aloc}

We give here a general proof by {\em reductio ad absurdum} that a simultaneous sharp localization of both ${\bm\psi}({\bm r})$ and ${\hat\chi}{\bm\psi}({\bm r})$ is not possible.

Take a function ${\bm\psi}_R({\bm r})$ sharply localized in $R$. Let us suppose that the function ${\hat\chi}{\bm\psi}_R({\bm r})$ is also sharply localized in some region $R'$. Then the sum of these two functions ${\bm\psi}_+({\bm r})=(1+{\hat\chi}){\bm\psi}_R({\bm r})$ is a function with compact support in the region $R\cup R'$. Due to the presence of the factor $(1+{\hat\chi})$, it is also an eigenfunction of the helicity operator belonging to the eigenvalue 1.  The Fourier transform of every integrable function with compact support is an entire function (see, for example, \cite{l}). The Fourier transform ${\tilde{\bm\psi}}_+({\bm k})$ of the function ${\bm\psi}_+({\bm r})$ is an eigenfunction of the helicity operator. Therefore, according to Eq.~(\ref{cp}), it must obey the relation
\begin{align}\label{contr}
{\bm k}\times{\tilde{\bm\psi}}_+({\bm k})=-i\sqrt{k_x^2+k_y^2+k_z^2}\;{\tilde{\bm\psi}}_+({\bm k}).
\end{align}
We arrived at a contradiction because the left hand side is an entire function but the right hand side is not because it contains a square root. Therefore, the function ${\hat\chi}{\bm\psi}_R({\bm r})$ cannot be sharply localized.

A less formal but more physical argument goes as follows. According to the formula (\ref{chi}), the function ${\hat\chi}{\bm\psi}_R({\bm r})$ has a $1/r^2$ tail extending to infinity. This fall-off is generic but we could obtain a faster decay by fine-tuning the function ${\bm\psi}_R({\bm r})$. Namely, we can require the vanishing of its successive moments. However, in this way we can never make the function ${\hat\chi}{\bm\psi}_R({\bm r})$ to vanish completely outside of $R$ because ${\bm\psi}_R({\bm r})$ cannot have all the moments equal to zero.

\section{}\label{adif}

We calculate the function ${\hat\chi}{\bm\varphi}_D$ using the expression for the convolution in terms of the Fourier transforms,
\begin{align}\label{conv}
\int\!d^3r'f({\bm r}-{\bm r}')g({\bm r}')=\int\!\frac{d^3k}{(2\pi)^3}e^{i{\bm k}\cdot{\bm r}}{\tilde f}({\bm k}){\tilde g}({\bm k}).
\end{align}
The Fourier representation of the function appearing in Eq.~(\ref{exp1}) has the form
\begin{align}\label{exp3}
\frac{{\rm Im}\,e^{-2\sqrt{1-ir/l}}}{r}&=\frac{1}{r\sqrt{\pi l}}\int_0^\infty\!\!dk\frac{\sin(kr)}{k^{3/2}}e^{-kl-1/kl}\nonumber\\
&=\frac{2\pi^{3/2}}{\sqrt{l}}\int\!\frac{d^3k}{(2\pi)^3}e^{i{\bm k}\cdot{\bm r}}\frac{e^{-kl-1/kl}}{k^{5/2}}.
\end{align}
In the first step we used the formula (Eq.~3.472.5. in \cite{gr})
\begin{align}\label{greq}
\int_0^\infty\!\!\frac{dx}{x^{n+1/2}}e^{-px-q/x}=(-1)^n\sqrt{\frac{\pi}{p}}\frac{\partial^n}{\partial q^n}e^{-2\sqrt{pq}}.
\end{align}
The imaginary parts in Eqs.~(\ref{exp3}) and (\ref{exp5}) can be extracted with the use of the following algebraic identity:
\begin{align}\label{ext}
\sqrt{2}\sqrt{a+ib}=\sqrt{|a+ib|+a}+\frac{ib}{|b|}\sqrt{|a+ib|-a}\,.
\end{align}

Putting together formulas (\ref{nonl}), (\ref{conv}), and (\ref{exp3}), we obtain
\begin{align}\label{exp4}
&{\hat\chi}{\bm\varphi}_D({\bm r})={\bm\nabla}\times\int\!d^3r'\frac{{\bm\varphi}_D({\bm r}')}{2\pi^2|{\bm r}-{\bm r}'|^2}\nonumber\\
&=\frac{2\pi^{3/2}}{\sqrt{l}}{\bm\nabla}\times\int\!\frac{d^3k}{(2\pi)^3}e^{i{\bm k}\cdot{\bm r}}\frac{e^{-kl-1/kl}}{k^{7/2}}{\bm m}\nonumber\\
&=\frac{1}{\sqrt{\pi l}}{\bm\nabla}\times\frac{1}{r}\int_0^\infty\!\!dk\frac{\sin(kr)}{k^{5/2}}e^{-kl-1/kl}{\bm m}\nonumber\\
&=\frac{1}{\sqrt{\pi l}}{\bm\nabla}\times{\rm Im}\frac{1}{r}\int_0^\infty\!\!dk\frac{e^{-k(l-ir)-1/kl}}{k^{5/2}}{\bm m}.
\end{align}
Using again Eq.~(\ref{greq}), we arrive at the final result
\begin{align}\label{exp5}
&{\hat\chi}{\bm\varphi}_D({\bm r})\nonumber\\
&=\frac{{\bm r}\times{\bm m}}{r^2}{\rm Im}\!\left[\left(i-\frac{l}{2r}-\frac{l}{r}\sqrt{1-\frac{ir}{l}}\!\right)\!e^{-2\sqrt{1-ir/l}}\right].
\end{align}


\begin{thebibliography}{2}
\bibitem{db} The denumerable dual bases introduced in \cite{sr} can be obtained by integrations with the sets of wavepackets.
\end{thebibliography}
\end{document}